\documentclass[10pt,twocolumn,showpacs,showkeys,preprintnumbers,amssymb,aps,superscriptaddress,pre]{revtex4-2}

\usepackage{amsmath, amssymb}
\usepackage{graphicx}
\usepackage{color,array}
\usepackage[colorlinks={true}]{hyperref}
\hypersetup{colorlinks=true,linkcolor=red,citecolor=blue,urlcolor=blue}
\usepackage{xcolor} 
\usepackage{color,array}
\usepackage{bm}
\usepackage{subfigure}
\usepackage{amsmath}
\usepackage{lipsum,appendix}
\usepackage{orcidlink}
\usepackage{pstricks}

\begin{document}


\title{Jahn-Teller Effect for Controlling Quantum Correlations in Hexanuclear Fe$^{3+}$ Magnets}

\author{Hamid Arian Zad~\!\!\orcidlink{0000-0002-1348-1777}}
\email{Corresponding author: hamid.arian.zad@upjs.sk}
\address{Department of Theoretical Physics and Astrophysics, Faculty of Science of P. J. \v{S}af{\'a}rik University, Park Angelinum 9, 040 01 Ko\v{s}ice, Slovak Republic}%

\author{Michal Ja{\v s}{\v c}ur~\!\!\orcidlink{0000-0003-0826-1961}}%
\address{Department of Theoretical Physics and Astrophysics, Faculty of Science of P. J. \v{S}af{\'a}rik University, Park Angelinum 9, 040 01 Ko\v{s}ice, Slovak Republic}%

\author{Asad Ali~\!\!\orcidlink{0000-0001-9243-417X}}%
\address{Qatar Centre for Quantum Computing, College of Science and Engineering, Hamad Bin Khalifa University, Doha, Qatar}%

\author{Saif Al-Kuwari~\!\!\orcidlink{0000-0002-4402-7710}}%
\address{Qatar Centre for Quantum Computing, College of Science and Engineering, Hamad Bin Khalifa University, Doha, Qatar}%

\author{Saeed Haddadi~\!\!\orcidlink{0000-0002-1596-0763}}%
\address{School of Particles and Accelerators, Institute for Research in Fundamental Sciences (IPM), P.O. Box 19395-5531, Tehran, Iran}%

\date{\today}

\begin{abstract}

We investigate the low-temperature magnetic and quantum properties of hexanuclear Fe\(^ {3+}_6\) complexes under an external magnetic field. We primarily study the impact of competing exchange interactions and their asymmetries induced by the Jahn-Teller distortion on the quantum properties of the complexes. The inequality in exchange interactions lifts the ground-state degeneracy that gives rise to complex quantum behavior. By constructing the ground-state phase diagram and analyzing magnetization, we identify key magnetic phases and critical phenomena. We further quantify quantum correlations using tripartite entanglement negativity and conditional von Neumann entropy to unveil how the Jahn-Teller effect enhances intra-triangle entanglement while modulating inter-triangle correlations. Our findings highlight the Fe\(^{3+}_6\) complex as a promising molecular platform for tunable quantum correlations, with potential applications in quantum information processing and molecular qubits.

\end{abstract}

\maketitle

{\it Introduction:} Molecular magnets \cite{Kahn, Chiesa2024, Chilton2013, Law2011, Schnack2005}, particularly transition-metal complexes with intrinsic quantum correlations, offer promising applications in quantum information science including spintronics \cite{Fursina2023,Kandala2017,Sanvito2011}, quantum communication \cite{Godfrin2017} and quantum computing \cite{Leuenberger2001,Pineda2021,Coronado2020}. This rapidly evolving field presents diverse opportunities by examining the unique magnetic and quantum properties of molecular systems \cite{Bode2023}. The study of these materials has gained significant attention due to their interdisciplinary relevance to bridge quantum information science, condensed matter physics, and coordination chemistry \cite{Carretta2021,Ollitrault2021,Chizzini2022, Lockyer2022,Soria2016}. Among them, hexanuclear iron(III) compounds \cite{Oyarzabal2015,Baca2013,Qian2008,Boudalis2006} exhibit a rich interplay of competing magnetic interactions, leading to unconventional quantum critical behaviors. A deeper understanding of these fundamental properties is essential for the development of next-generation molecular-based quantum technologies.

However, under the influence of the Jahn-Teller (JT) effect \cite{Boudalis2006,Jahn1937,Koppel2009,Halcrow2013}, these compounds undergo a spontaneous symmetry-breaking distortion that lifts the degeneracy of their electronic states \cite{Boudalis2006}. As predicted by Jahn and Teller in 1937, a non-linear system in a degenerate electronic state is inherently unstable and will distort to remove the degeneracy, thereby its total energy is lowered \cite{Jahn1937}. In the case of Fe\(^{3+}_6\) complexes, the JT effect alters the geometric and electronic structure of the Fe\(_3\)O subunits, subtly modifying the exchange interactions and consequently reshaping the magnetic ground state. This distortion introduces anisotropy into the system, affecting both the intra- and inter-triangle couplings and influencing the quantum correlations and spin frustration within the molecule. Hence, the JT effect plays a crucial role in determining the fine structure of the energy landscape, where it leads to distinct magnetic behavior that would not be observed in an idealized symmetric model with equilateral triangular subunits.

When an external field is applied, the competition between exchange interactions leads to intriguing quantum phenomena, including field-induced phase transitions and entanglement structures that remain robust at low temperatures.
Here, we systematically investigate the ground-state phase diagram and low-temperature quantum properties of the isotropic XXX Heisenberg model on a hexanuclear spin-1/2 system possessing three different exchange couplings that characterize the Fe\(^{3+}_6\) complexes such as $\text{Fe}^{3+}_6(\mu_3\text{-}\text{O})_2(\mu\text{-}\text{OH})_2\{\mu\text{-}(\text{C}_6\text{H}_{11})_2\text{PO}_2\}_6(\mu\text{-}t\text{BuCO}_2)_6\\
(\eta^1\text{-}\text{OH}_2)_2 \textperiodcentered 2\text{CH}_3\text{CN} \textperiodcentered \text{CH}_2\text{Cl}_2$ \cite{Oyarzabal2015}.
A key feature of hexanuclear Fe\(^{3+}_6\) complexes is the presence of multiple exchange interactions, which determine the ground-state spin configuration and influence the response of the system to external perturbations such as magnetic field \cite{Oyarzabal2015,Baca2013,Qian2008,Boudalis2006}.
In particular, the Fe\(^{3+}_6\) molecular structure comprises two almost identical Fe\(_3\)O triangular subunits that exhibit a quantum ground state with \(S=1/2\) per triangular subunit.
These iron-based molecular systems exhibit competing exchange interactions, \( J_1 \), \( J_2 \), and \( J_3 \) under the influence of the JT effect, leading to \( S = \frac{1}{2} \) ground states for both \(\text{Fe}_3\text{O}\) triangular subunits. The inequality of \( J_{ij} \) lifts the degeneracy of the ground states (\( S \geq \frac{1}{2} \)) that is a consequence of magnetic JT distortion. The two triangular subunits are coupled via the inter-triangle interaction \( J_3 \), which plays a crucial role in the quantumness of the system, particularly when this interaction is non-negligible \cite{Oyarzabal2015}.

In this letter, we construct the ground-state phase diagram in the plane of external magnetic field \(B\), intra-triangle exchange coupling \(J_2\) and inter-triangle exchange coupling \(J_3\). Then, we analyze the magnetization process and explore how the entanglement structure evolves with varying field strengths under JT distortion. Yo quantify quantum entanglement \cite{Horodecki2009} within the subunits we use the tripartite entanglement negativity \cite{Vargova2023,ArianZad2025,Ghannadan2025}, a concept introduced by Peres and Horodecki \cite{Peres1996,Horodecki1996,Vidal2002}. We also examine the conditional von Neumann entropy \cite{Cerf1997,Lami2017,Hirche2020,Schwonnek2024}, a key measure that captures quantum correlations between different Fe\(_3\)O triangular subunits. This approach provides a comprehensive overivew of how quantum information is distributed across the molecular system.

Our findings contribute to the broader understanding of entanglement in molecular magnets at low temperature and highlight the potential of hexanuclear Fe\(^{3+}\) complexes as testbeds for quantum correlations in condensed matter systems. By unveiling the fundamental entanglement mechanisms in these systems, this work opens new avenues for their potential applications in emerging quantum technologies.


The model depicted in Fig. \ref{fig:model} consists of two identical isosceles Fe\(_3(\mu_3\)-O) triangles. The magnetic interactions between the spin-only Fe\({}^{3+}\) ions (with \(S = \frac{1}{2}\)) are governed by three exchange coupling constants (\(J_1\), \(J_2\), and \(J_3\)). These interactions can be described by the following Hamiltonian:
\begin{figure}[tbp]
	\centering
	\includegraphics[scale=0.3,trim=50 50 10 00, clip]{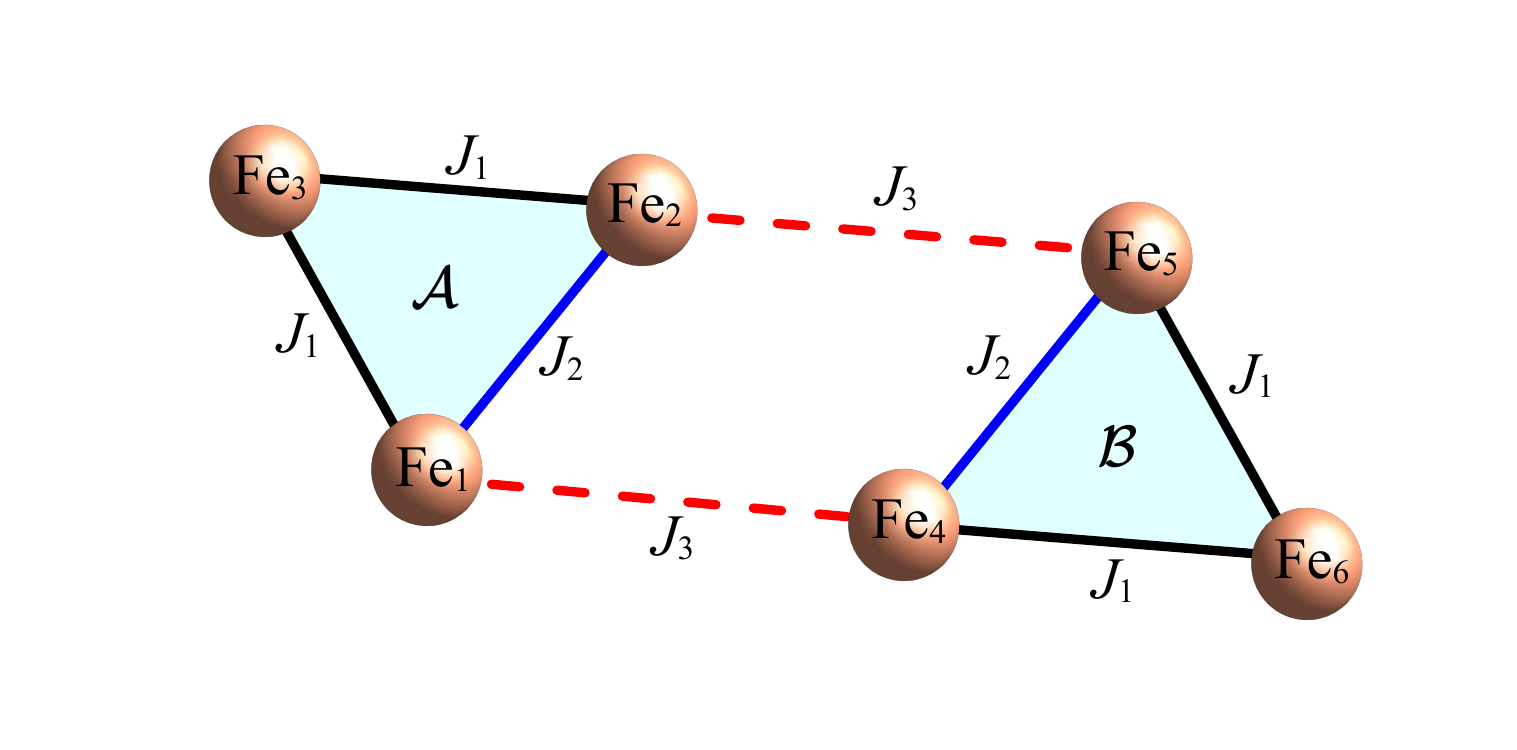} %
	\vspace{-0.25cm}
	\caption{The molecular structure of the Fe$^{3+}_6$ complex. We label the first triangular subunit as subsystem $\mathcal{A}$ and the second one as subsystem $\mathcal{B}$.}
	\label{fig:model}
\end{figure}

\begin{eqnarray}\label{Eq:Hamiltonian}
H &=& J_1 (\hat{\mathbf{S}}_1 \cdot \hat{\mathbf{S}}_3
+ \hat{\mathbf{S}}_2 \cdot \hat{\mathbf{S}}_3
+ \hat{\mathbf{S}}_4 \cdot \hat{\mathbf{S}}_6
+ \hat{\mathbf{S}}_5 \cdot \hat{\mathbf{S}}_6 ) \nonumber \\
&&
+ J_2 (\hat{\mathbf{S}}_1 \cdot \hat{\mathbf{S}}_2
+ \hat{\mathbf{S}}_4 \cdot \hat{\mathbf{S}}_5 )
+ J_3 (\hat{\mathbf{S}}_1 \cdot \hat{\mathbf{S}}_4
+ \hat{\mathbf{S}}_2 \cdot \hat{\mathbf{S}}_5 )  \nonumber \\
	&&- g\mu_\text{B} B\sum_{i=1}^6\hat{S}_i^z,
	\label{H}
\end{eqnarray}
This Hamiltonian represents six spin-1/2 particles and three different isotropic Heisenberg exchange interactions.
Here, $\hat{S}^{\alpha}_j$ ($\alpha =x,y,z$) are the spin-1/2 operators assigned to the $\text{Fe}^{3+}$ ions.
Fitting performed between theoretical analysis and the experimental data of the magnetization and susceptibility constrains the three exchange constants (\(J_1, J_2, J_3\)) to maintain their antiferromagnetically relative order \(J_1 > J_2 > J_3\). The resulting fits show an agreement with the experimental data.
 Here, $g = 2.0$ is the gyromagnetic ratio, $\mu_\text{B}$ is Bohr magneton and $B$ is the magnetic field applied along the $z$-direction. According to the experimental analysis of the $\text{Fe}^{3+}$ complex reported in \cite{Oyarzabal2015}, the two triangular units are not equilateral, most probably due to the JT effect. The three distinct coupling constants in this iron compound are found to be antiferromagnetic, with values $J_1 = 42.2 \, \text{cm}^{-1}$, $J_2 = 34.2 \, \text{cm}^{-1}$, and $J_3 = 0.015 \, \text{cm}^{-1}$.

The partition function $Z = \sum_i e^{-\beta E_i}$ of the system is given by the summation over the Boltzmann-weighted eigenenergies \( E_i \) of the Hamiltonian (\ref{Eq:Hamiltonian}), where \( \beta = \frac{1}{k_\text{B} T} \), with \( k_\text{B} \) being the Boltzmann constant and \( T \) the temperature. The Gibbs free energy follows from the partition function as $G = -k_\text{B} T \ln Z$. From this, the magnetization can be determined via
$M = -\left( \frac{\partial G}{\partial B} \right)_T$.

\begin{figure*}[tbp]
	\centering
	\includegraphics[scale=0.3,trim=100 50 00 50, clip]{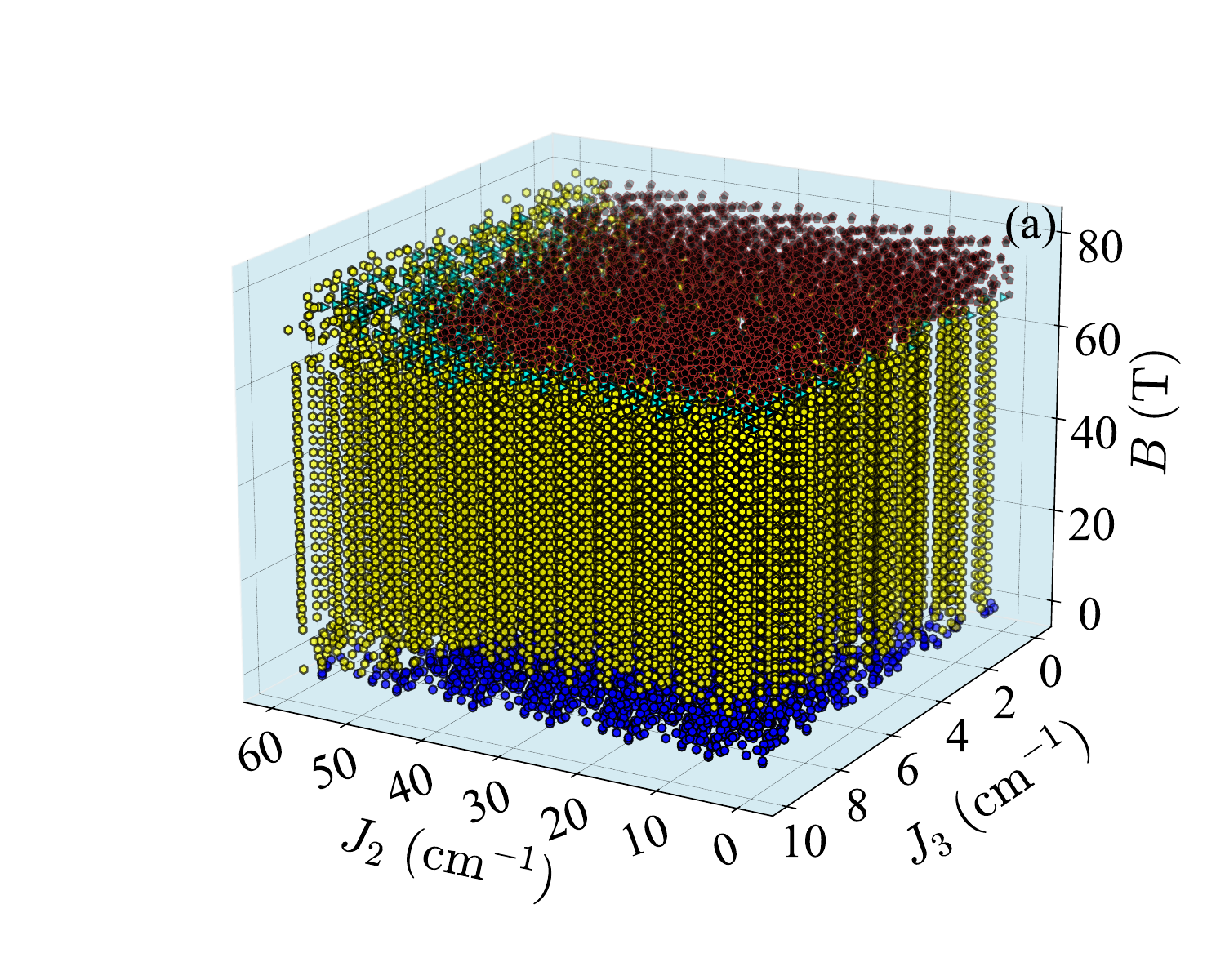} %
	\includegraphics[scale=0.25,trim=10 00 00 50, clip]{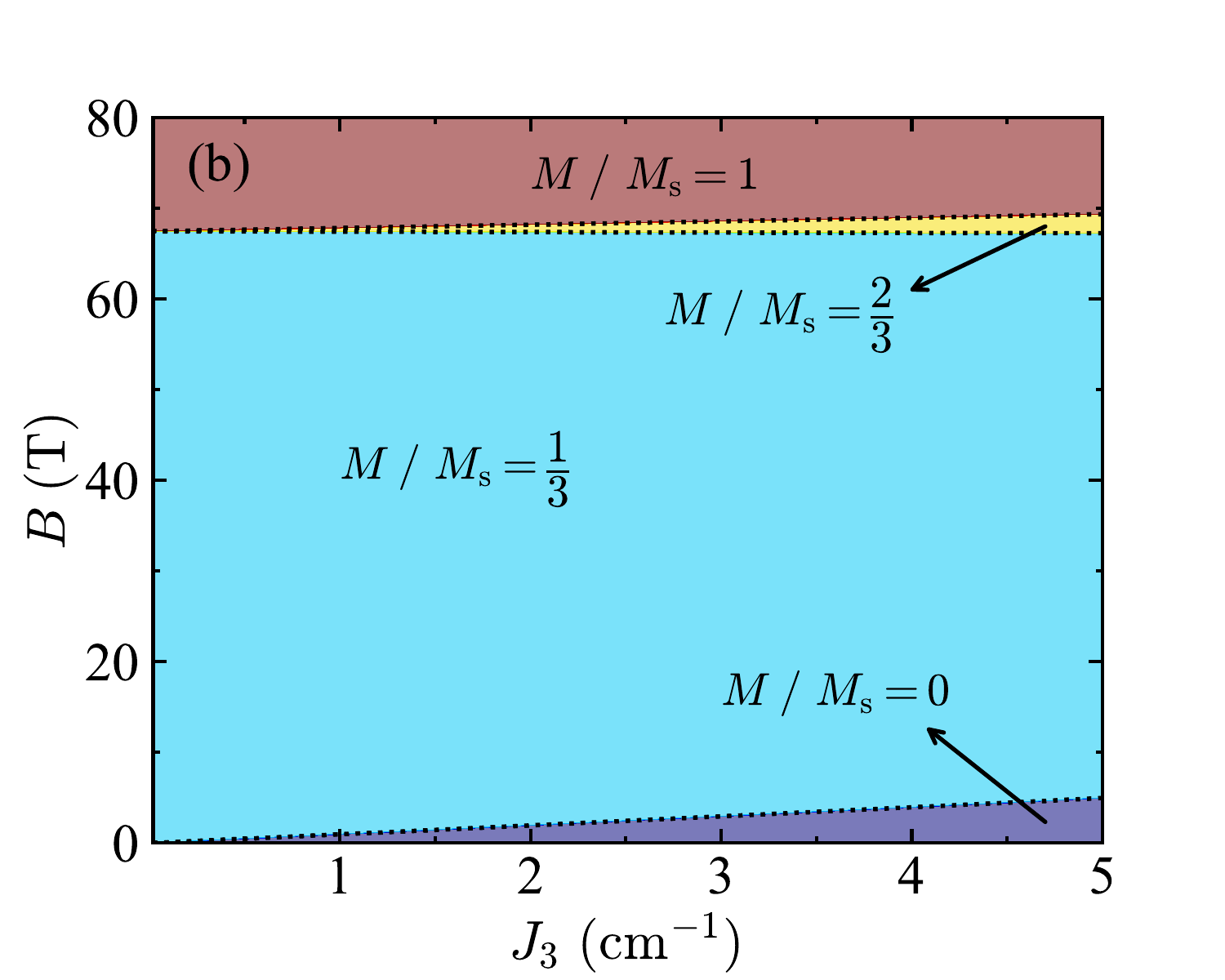} %
	\includegraphics[scale=0.25,trim=00 00 00 50, clip]{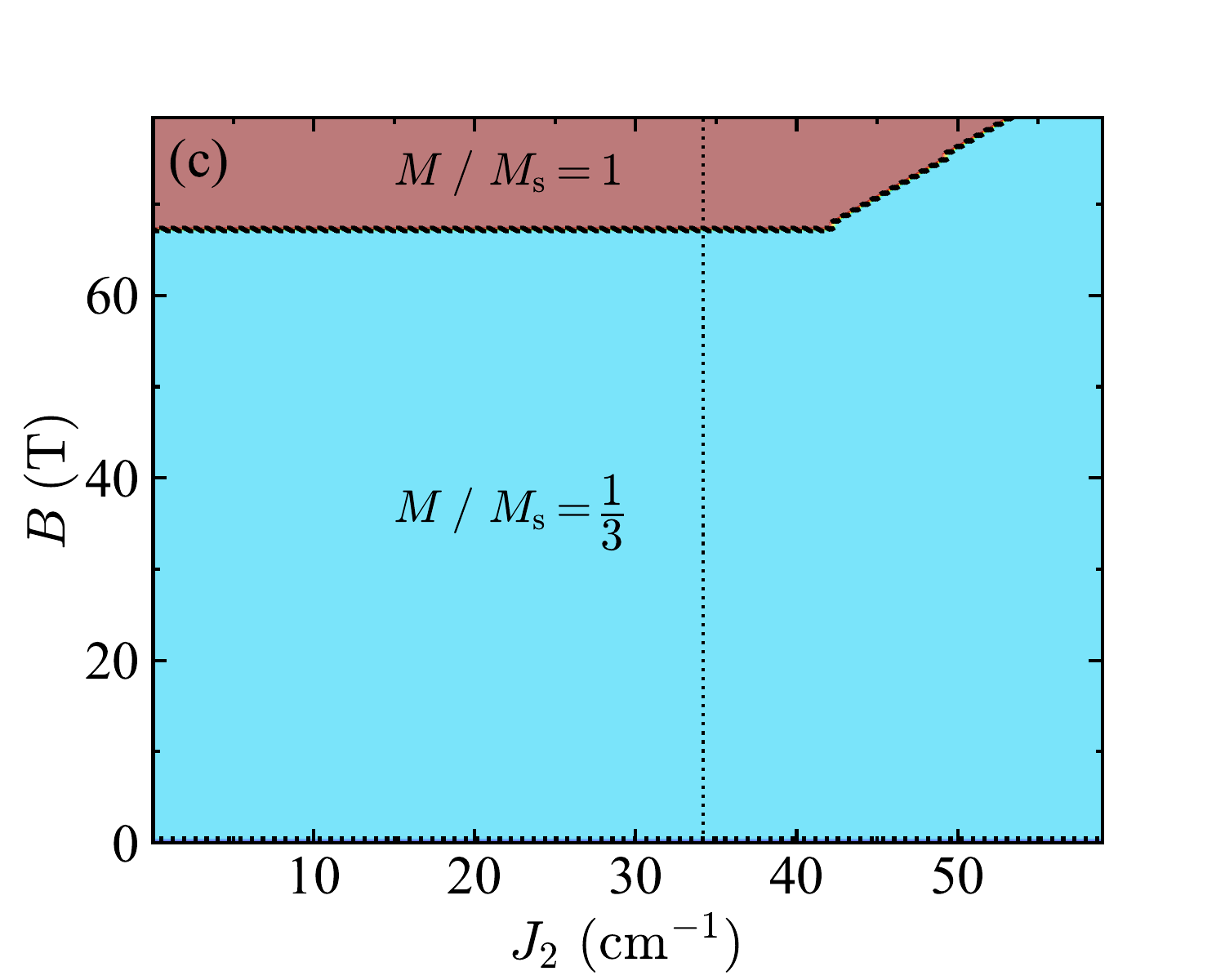} %
	\includegraphics[scale=0.25,trim=00 00 00 50, clip]{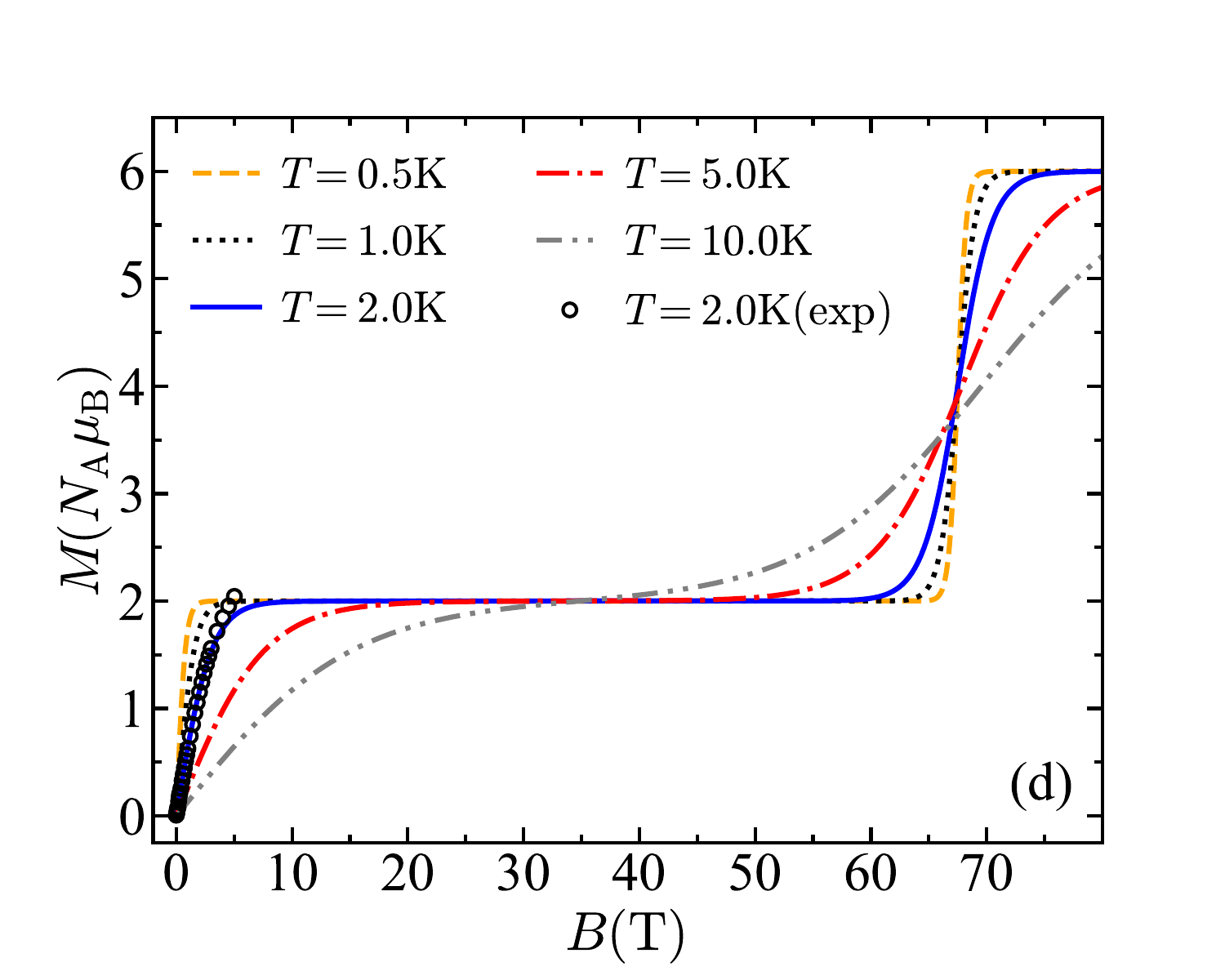} %
	\vspace{-0.25cm}
	\caption{	
		(a) Ground-state phase diagram of the $\text{Fe}^{3+}$ complex described by the Hamiltonian (\ref{Eq:Hamiltonian}) in the ($J_2, J_3, B$) parameter space, assuming $J_1 = 42.2\, \text{cm}^{-1}$. The different ground states are represented by distinct symbols: black pentagons for $M/M_\text{s} = 1$,  cyan triangles for $M/M_\text{s} = \frac{2}{3}$, yellow hexagons for $M/M_\text{s} = \frac{1}{3}$, and blue circles for $M/M_\text{s} = 0$.
		(b) Ground-state phase diagram in the $B-J_3$ plane for fixed values of $J_1 = 42.2\, \text{cm}^{-1}$ and $J_2 = 34.2\, \text{cm}^{-1}$.
		(c) Ground-state phase diagram in the $B-J_2$ plane, assuming $J_1 = 42.2\, \text{cm}^{-1}$ and $J_3 = 0.015\, \text{cm}^{-1}$.
		(d) Magnetization as a function of the magnetic field $B$ for various fixed temperatures $T$. Experimental data from ~\cite{Oyarzabal2015} are shown as black circles, while theoretical magnetization curves are represented by lines. The blue solid line corresponds to the best-fit magnetization curve, obtained using the parameter set $J_1 = 42.2\, \text{cm}^{-1}$, $J_2 = 34.2\, \text{cm}^{-1}$, and $J_3 = 0.015\, \text{cm}^{-1}$, as reported in ~\cite{Oyarzabal2015}.}
	\label{fig:GSPD_Mag_BJ2J3}
\end{figure*}

We present the ground-state phase diagram of the model described by the Hamiltonian (\ref{Eq:Hamiltonian}) in Fig.~\ref{fig:GSPD_Mag_BJ2J3}(a) within the three-dimensional ($J_2, J_3, B$) parameter space.
The diagram reveals four distinct regions, each marked by different symbols corresponding to the ground states with magnetization values $M/M_\text{s} = 0$, $\frac{1}{3}$, $\frac{2}{3}$, and $1$. From this figure, it is evident that the phase with $M/M_\text{s} = \frac{1}{3}$ is the most dominant across the parameter space. To further illustrate this, Fig.~\ref{fig:GSPD_Mag_BJ2J3}(b) displays the ground-state phase diagram in the $B$-$J_3$ plane for fixed values of $J_1 = 42.2\, \text{cm}^{-1}$ and $J_2 = 34.2\,\text{cm}^{-1}$. As the inter-triangle interaction $J_3$ increases, two narrow phase regions emerge, corresponding to the zero-magnetization phase and the intermediate plateau at $M/M_\text{s} = \frac{2}{3}$, as indicated by the cyan triangle symbols in Fig.~\ref{fig:GSPD_Mag_BJ2J3}(a).

To investigate the influence of JT distortion on the phase boundaries, we present in Fig.~\ref{fig:GSPD_Mag_BJ2J3}(c) the ground-state phase diagram in the $B$-$J_2$ plane for the fixed $J_1 = 42.2\, \text{cm}^{-1}$ and $J_3 = 0.015\,\text{cm}^{-1}$. It should be noted that below $J_2 = J_1$, where the JT distortion is present, the extent of the $M/M_\text{s} = \frac{1}{3}$ phase remains unchanged. However, for $J_2 > J_1$, the phase boundary of the $M/M_\text{s} = \frac{1}{3}$ state expands, indicating an increased stability of this phase.

In order to verify the theoretical phase diagram through magnetization behavior, Fig.~\ref{fig:GSPD_Mag_BJ2J3}(d) depicts the magnetization of the Fe$^{3+}$ complex as a function of the magnetic field at various temperatures, assuming $J_1 = 42.2\, \text{cm}^{-1}$, $J_2 = 34.2\,\text{cm}^{-1}$ and $J_3 = 0.015\,\text{cm}^{-1}$. The low-temperature magnetization curve (orange dashed line) in Fig.~\ref{fig:GSPD_Mag_BJ2J3}(d) exhibits a broad intermediate plateau at $M/M_\text{s} = \frac{1}{3}$, which aligns with the vertical dotted line in Fig.~\ref{fig:GSPD_Mag_BJ2J3}(c). As the temperature increases, this plateau gradually vanishes due to the quantum superposition of other magnetic states. Furthermore, the best fit of the theoretical results (blue solid line) with the experimental magnetization data for the Fe$^{3+}$ compound at $2\,\text{K}$ (black circles) up to $5\,\text{T}$ is shown, reinforcing the validity of our modelling.

\label{Sec:entanglement}
To quantify the degree of quantum correlation between all entities of a single triangular subunit ($\mathcal{A}$ or $\mathcal{B}$) of the hexanuclear Fe$^{3+}$ system, we employ the tripartite entanglement negativity $\mathcal{N}_{123}$ \cite{Vargova2023,ArianZad2025,Ghannadan2025}. This measure is defined as the geometric mean of three bipartite negativities:

\begin{equation}\label{Eq:triNeg}
	\mathcal{N}_{123} = \sqrt[\raisebox{0.8ex}{\small $\frac{1}{3}$}]{\mathcal{N}_\mathrm{1|23}\,\mathcal{N}_\mathrm{2|13}\,\mathcal{N}_\mathrm{3|12}}.
\end{equation}

The bipartite negativities \(\mathcal{N}_{1|23}\), \(\mathcal{N}_{2|13}\), and \(\mathcal{N}_{3|12}\) are determined by tracing out the degrees of freedom associated with the subsystem $\mathcal{B}$, thereby reducing the system to a the subsystem $\mathcal{A}$ including Fe$_1$, Fe$_2$ and Fe$_3$ (see Fig. \ref{fig:model}). Each bipartite negativity quantifies the entanglement between an individual Fe$^{3+}$ ion and the remaining two Fe$^{3+}$ ions.
This type of entanglement can be computed using the approach introduced by Vidal and Werner in \cite{Vidal2002}, which defines bipartite negativity as the sum of the absolute values of all negative eigenvalues of the partially transposed density matrix \(\hat{\rho}_{123}^{\text{T}_\text{a}}\), where the partial transposition is performed over subsubsystem ``a" (Fe$_1$, Fe$_2$ or Fe$_3$). For example, this can be expressed as
$\mathcal{N}_{1|23} = \sum_{(\lambda_{1|23})_i < 0} |\lambda_{1|23}|$, where \(\lambda\) denotes the eigenvalues of the reduced density matrix \(\rho_{123}^{\text{T}_1}\) that is partially transposed with respect to Fe$_1$. The reduced density matrices and their transpose along with the eigenvalues $\lambda_i$ are obtained using exact numerical methods implemented in the {\it QuTip} package \cite{Johansson2012}.

\begin{figure}[tbp]
	\centering
	\includegraphics[scale=0.31,trim=10 00 00 20, clip]{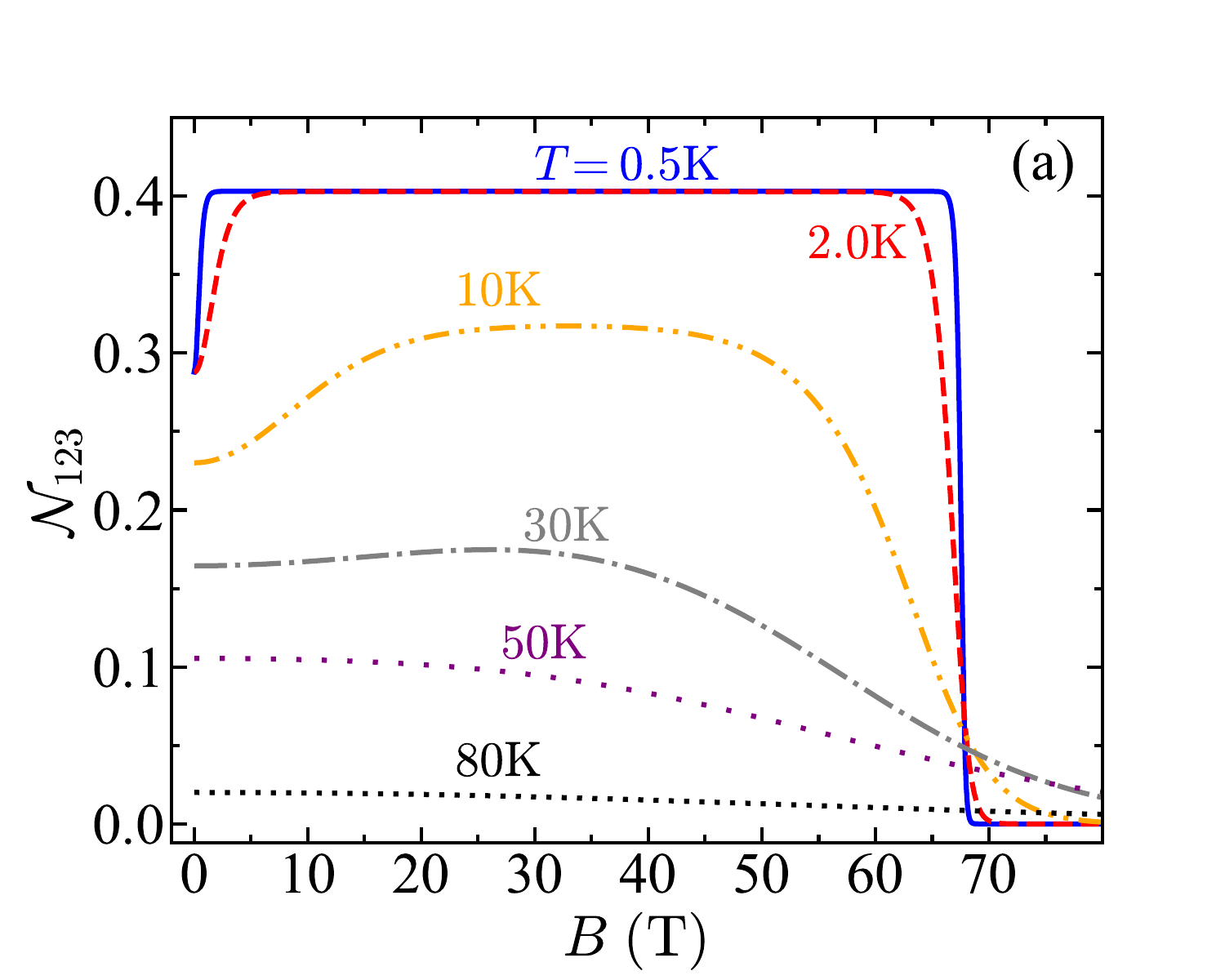} %
	\includegraphics[scale=0.31,trim=10 00 00 20, clip]{ 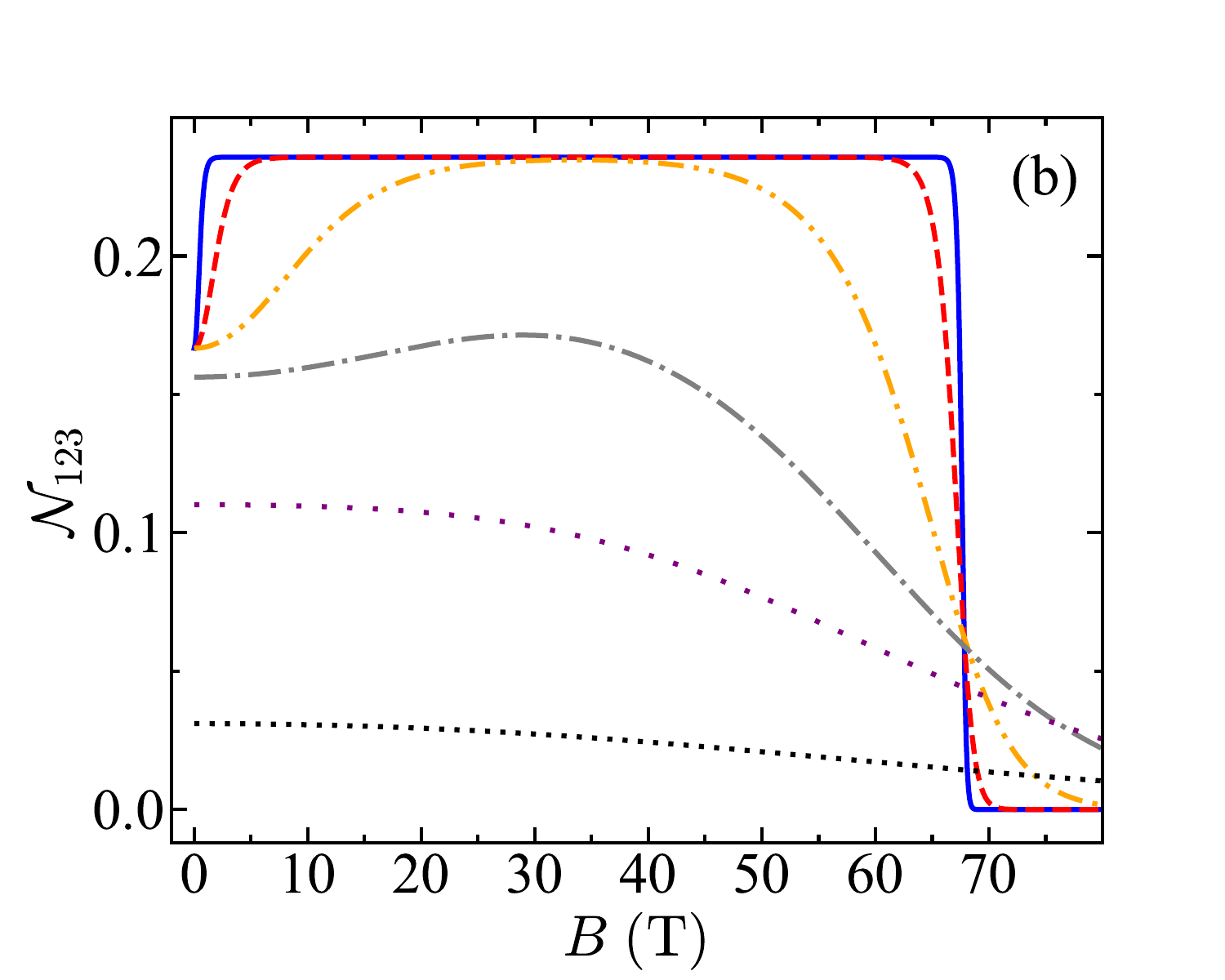}
	\vspace{-0.2cm}
	\caption{(a) Magnetic-field dependence of the tripartite negativity $\mathcal{N}_{123}$ of the Fe$^{3+}$ complex under JT effect at different temperatures.  The coupling constants $J_1 = 42.2\, \text{cm}^{-1}$, $J_2 = 34.2\,\text{cm}^{-1}$, $J_3 = 0.015\,\text{cm}^{-1}$, and the gyromagnetic factor $g = 2.0$ are taken from previous experimental analysis \cite{Oyarzabal2015}.
	 (b) The negativity $\mathcal{N}_{123}$ with equilateral triangle shape of subunits with  $J_1 = J_2 = 42.2\, \text{cm}^{-1}$ and $J_3 = 0.015\,\text{cm}^{-1}$.}
	\label{fig:Neg_BT}
\end{figure}

Figure \ref{fig:Neg_BT}(a) shows the genuine tripartite negativity \(\mathcal{N}_{{123}}\) in the triangular subsystem $\mathcal{A}$  under the influence of the JT effect, plotted as a function of the magnetic field at various fixed temperatures. This analysis is conducted using the exchange interaction parameters \(J_1 = 42.2\, \text{cm}^{-1}\), \(J_2 = 34.2\,\text{cm}^{-1}\), and \(J_3 = 0.015\,\text{cm}^{-1}\), where the JT distortion is reflected in the condition \(J_1 > J_2\).
By inspecting this figure, it is evident that as the magnetic field increases, the tripartite negativity initially rises from a nonzero value before stabilizing into a broad, flat plateau at \(\mathcal{N}_{123} \approx 0.4\). This plateau corresponds to the ground state characterized by \(M/M_\text{s} = \frac{1}{3}\). However, as the field approaches the transition point near \(B \approx 68\,\text{T}\), the tripartite negativity undergoes a sharp decline, indicating a transition to a fully polarized state.
As the temperature increases, this distinct stepwise behavior gradually fades, leading to a smooth, monotonic decrease in tripartite negativity with increasing magnetic field.

To further investigate the role of the JT effect on tripartite entanglement, we plot in Fig. \ref{fig:Neg_BT}(b) the tripartite negativity \(\mathcal{N}_{123}\) as a function of the magnetic field for the case where both triangular subunits $\mathcal{A}$ and $\mathcal{B}$ are equilateral (\(J_1 = J_2\)). It is evident that under this condition, the degree of entanglement significantly decreases, reaching a value of \(\mathcal{N}_{123} \approx 0.25\), which is approximately half of the value observed in the presence of the JT effect (see Fig. \ref{fig:Neg_BT}(a)). This finding suggests that the JT effect can substantially enhance entanglement between the spins of the subunits of the hexanuclear Fe\(^ {3+}\) complexes, potentially playing a crucial role in optimizing quantum correlations within such molecular systems.

\begin{figure*}[tbp]
	\centering
	\includegraphics[scale=0.24,trim=10 00 20 20, clip]{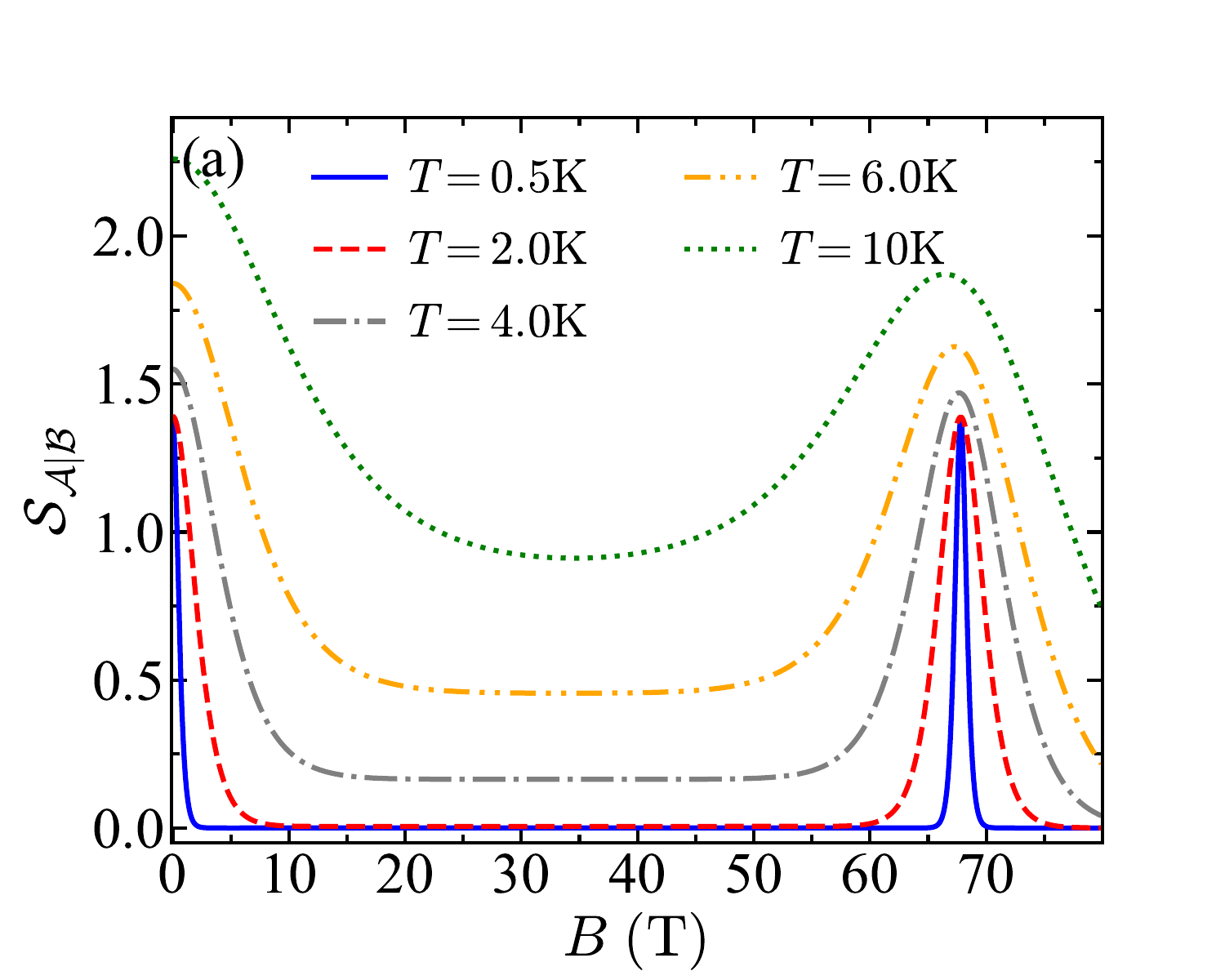}
	\includegraphics[scale=0.24,trim=10 00 20 20, clip]{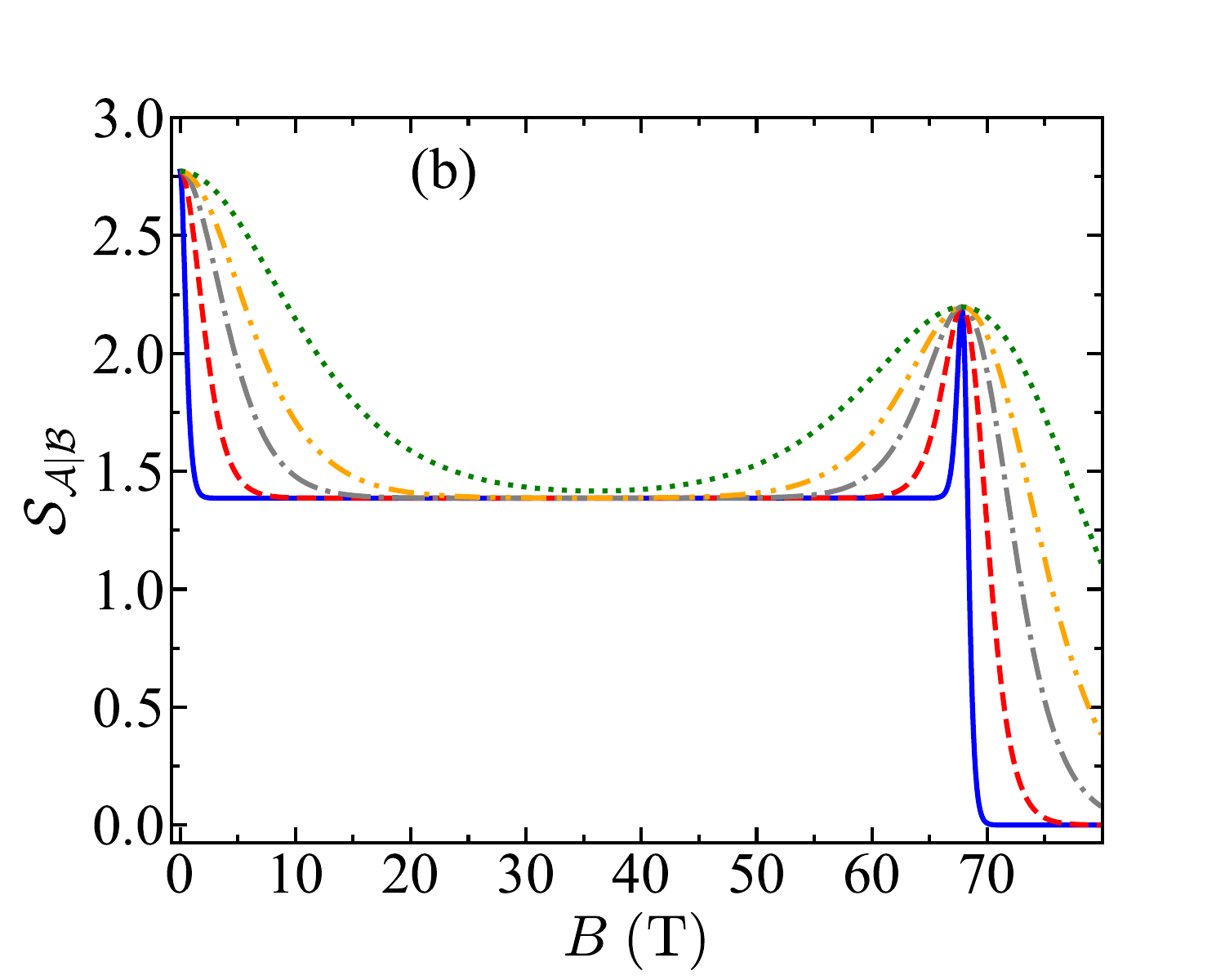}
	\includegraphics[scale=0.24,trim=10 00 20 20, clip]{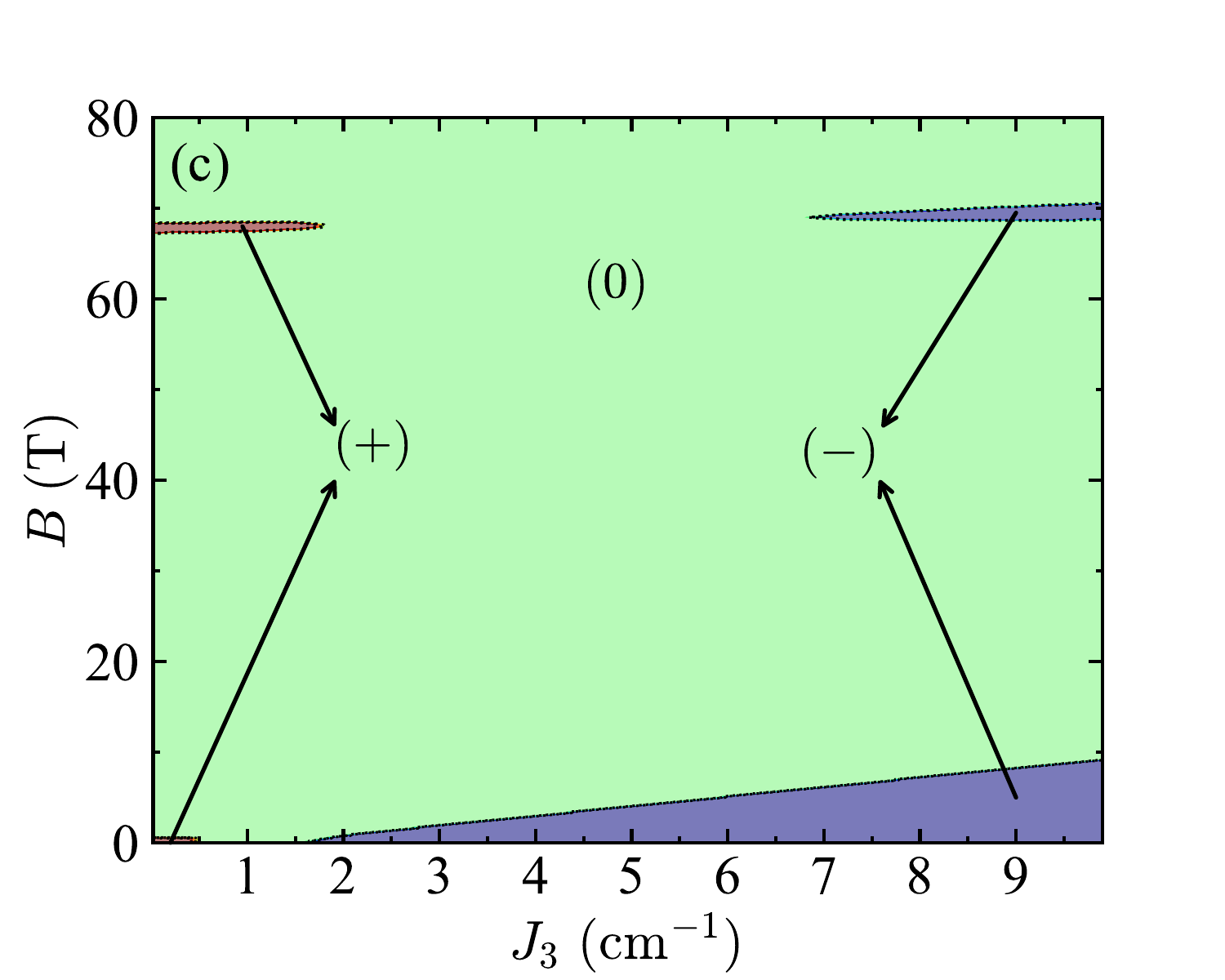}
	\vspace{-0.2cm}
	\caption{(a) Magnetic-field dependence of the conditional entropy $\mathcal{S}_\mathcal{A|B}$ of the Fe$^{3+}$ complex under JT effect at different temperatures. The coupling constants are $J_1 = 42.2\, \text{cm}^{-1}$, $J_2 = 34.2\,\text{cm}^{-1}$, $J_3 = 0.015\,\text{cm}^{-1}$, and the gyromagnetic factor is $g = 2.0$  \cite{Oyarzabal2015}.
	(b) The conditional entropy $\mathcal{S}_\mathcal{A|B}$ of the equilateral triangle shape of the subsystems $\mathcal{A}$ and $\mathcal{B}$ with $J_1 = J_2 = 42.2\, \text{cm}^{-1}$ and $J_3 = 0.015\,\text{cm}^{-1}$.
	(c) Conditional von Neumann entropy $\mathcal{S}_\mathcal{A|B}$ in the ($B, J_3$) plane, assuming $J_1 = 42.2\, \text{cm}^{-1}$ and $J_2 = 34.2\,\text{cm}^{-1}$.}
	\label{fig:condEnt_BT}
\end{figure*}

We present in Fig. \ref{fig:condEnt_BT}(a) the magnetic field dependence of the conditional von Neumann entropy \(\mathcal{S}_\mathcal{A|B}=\mathcal{S}_\mathcal{AB}-\mathcal{S}_\mathcal{B}\) at various fixed temperatures to further explore the impact of JT distortion on the degree of quantum correlation between the two triangular subunits $\mathcal{A}$ and $\mathcal{B}$ of the hexanuclear Fe\(^ {3+}\) complex. This analysis is based on the exchange interaction parameters \(J_1 = 42.2\, \text{cm}^{-1}\), \(J_2 = 34.2\,\text{cm}^{-1}\), and \(J_3 = 0.015\,\text{cm}^{-1}\) \cite{Oyarzabal2015}.
Abrupt changes in the conditional entropy occur near the transition fields, which can be attributed to quantum superposition among multiple phases. This serves as a strong indication of quantum correlations between the two triangular subunits. At low temperatures and within the moderate field range \(5\,\text{T} < B < 65\,\text{T}\), the conditional entropy remains zero, signifying the absence of inter-subunit correlations. However, as the temperature increases, \(\mathcal{S}_\mathcal{A|B}\) also increases, reflecting the strengthening of quantum correlations between subunits $\mathcal{A}$ and $\mathcal{B}$.
Figure \ref{fig:condEnt_BT}(b) illustrates the conditional von Neumann entropy \(\mathcal{S}_\mathcal{A|B}\) for the case where the triangular subunits are equilateral, i.e. \(J_1 = J_2 = 42.2\, \text{cm}^{-1}\). Comparing Figs. \ref{fig:condEnt_BT}(a) and \ref{fig:condEnt_BT}(b), we observe that the entropy \(\mathcal{S}_\mathcal{A|B}\), which quantifies inter-triangle quantum correlations, is higher when \(J_1 = J_2\). This suggests that JT distortion enhances the degree of tripartite entanglement within each triangular subunit, while simultaneously reducing quantum correlations between the two triangular subunits. This characteristic of the JT effect is particularly advantageous for quantum information processing, as it facilitates the development of molecular qubits for quantum computing applications.

Figure \ref{fig:condEnt_BT}(c) shows the conditional von Neumann entropy \(\mathcal{S}_\mathcal{A|B}\) in the \(B-J_3\) plane, assuming \(J_1 = 42.2\, \text{cm}^{-1}\) and \(J_2 = 34.2\,\text{cm}^{-1}\). Clearly, the inter-triangle exchange interaction \(J_3\) plays a crucial role in shaping the nature of quantum correlations between the two subunits. In particular, stronger \(J_3\) leads to negative values of the conditional entropy \(\mathcal{S}_\mathcal{A|B}\), indicating that the two triangular subunits become entangled \cite{Horodecki2005}.


In this letter, we investigated the ground-state phase diagram and quantum correlations in the hexanuclear Fe\(^ {3+}\) complex under the influence of the JT effect. Our findings indicate that the magnetic phase with one-third magnetization plateau dominates across the parameter space. This state remains stable in the presence of the JT effect characterized by \( J_1 > J_2 \).
The role of JT effect in modifying tripartite entanglement within a triangular subunit has been studied. Our results demonstrate that the JT distortion enhances multipartite entanglement in the subunits. When the JT effect is absent and the triangular units are equilateral, tripartite entanglement is significantly reduced, confirming that the JT distortion plays a key role in strengthening quantum correlations at the molecular level.
Moreover, our analysis of the conditional von Neumann entropy demonstrated that the JT effect may diminish inter-triangle quantum correlations. At low temperatures, the two triangular subunits remain uncorrelated in a moderate field range, but increasing temperature leads to a rise in conditional entropy, which suggests enhanced quantum correlations. We have found that stronger inter-triangle interaction leads to negative conditional entropy, signifying entanglement between the two subunits.

Overall, our study demonstrates that JT effect increases intra-triangle tripartite entanglement, but may decrease the inter-triangle correlations. These findings suggest that JT distortions could be harnessed for optimizing molecular qubits in quantum information processing. By controlling exchange interactions and external magnetic fields, Fe\(^ {3+}\) complexes may serve as promising candidates for designing quantum computing architectures.


{\it Acknowledgments --} H.A.Z. acknowledges the financial support provided under the postdoctoral fellowship program of P. J. \v{S}af\'{a}rik University in Ko\v{s}ice, Slovakia. This research was funded by Slovak Research and Development Agency under the contract No. APVV-20-0150 and  The Ministry of Education, Research, Development and Youth of the Slovak Republic under the grant number VEGA 1/0695/23. S.H. thanks the School of Particles and Accelerators at the Institute for Research in Fundamental Sciences for their financial support.

\end{document}